\documentclass[runningheads]{llncs}
\usepackage[T1]{fontenc}
\usepackage{graphicx}
\usepackage{pgfplots}
\usepackage{multirow, array}

\pgfplotsset{compat=1.18}
\begin{document}

\title{Multi-Resolution Fusion for Fully Automatic Cephalometric Landmark Detection}
\titlerunning{Multi-Resolution Fusion for Fully Automatic CL-Detection}
\author{Dongqian Guo \and Wencheng Han}
\authorrunning{D. Guo et al.}
\institute{SKL-IOTSC, University of Macau, Taipa, Macau \\
    \email{mike.guodongqian@connect.umac.mo, wencheng256@gmail.com}
}

\maketitle

\begin{abstract}
Cephalometric landmark detection on lateral skull X-ray images plays a crucial role in the diagnosis of certain dental diseases. Accurate and effective identification of these landmarks presents a significant challenge. Based on extensive data observations and quantitative analyses, we discovered that visual features from different receptive fields affect the detection accuracy of various landmarks differently. As a result, we employed an image pyramid structure, integrating multiple resolutions as input to train a series of models with different receptive fields, aiming to achieve the optimal feature combination for each landmark. Moreover, we applied several data augmentation techniques during training to enhance the model's robustness across various devices and measurement alternatives. We implemented this method in the Cephalometric Landmark Detection in Lateral X-ray Images 2023 Challenge and achieved a Mean Radial Error (MRE) of 1.62 mm and a Success Detection Rate (SDR) 2.0mm of 74.18\% in the final testing phase.

\end{abstract}

\section{Introduction}
Cephalometric analysis plays a pivotal role in the diagnosis and treatment of dental diseases. This procedure allows clinicians to observe the intricacies of the skull and both the upper and lower jaw structures, facilitating the swift and accurate identification of relevant pathological changes. Typically, this analysis is performed on a patient's lateral skull X-ray image, where precise recognition of anatomical landmarks is essential for further scrutiny. In the past, these landmarks were manually annotated by clinical doctors. This process was not only time-consuming and labor-intensive but also heavily reliant on a doctor's clinical experience and subjective judgment. Consequently, this made it susceptible to errors, potentially hindering subsequent diagnoses and treatments. As a result, the introduction of semi-automatic or fully automatic methods for skull landmark identification has become imperative.

At the inception of the problem, researchers attempted to utilize computer programs to simulate manual annotation rules for automating the task \cite{levy1986knowledge}. However, as the complexity of the images increased, the rules used by clinical doctors for annotating landmarks became increasingly difficult to quantify. Consequently, this method soon became obsolete. Subsequently, some proposed algorithms based on pattern matching \cite{cardillo1994image,el2004automatic}. However, due to the vast variations in skull shapes and the features surrounding the landmarks among different individuals, these methods' sensitivity to individual differences rendered them incapable of achieving satisfactory recognition accuracy. Later on, machine learning methods began to gain popularity. In 2014 and 2015, the IEEE ISBI hosted a cephalometric landmark detection challenge \cite{wang2015evaluation,wang2016benchmark}, and a review article was written based on some of the state-of-the-art methods at that time. The top-performing methods in the competition were based on random forests \cite{ibragimov2014automatic,lindner2016fully,tim2015fully}. In recent years, with the advancement of deep learning \cite{lecun2015deep}, this technology has been extensively applied to the field of medical image analysis \cite{arik2017fully,kim2020web,payer2019integrating}. A study based on "Attentive Feature Pyramid Fusion and Regression-Voting" \cite{chen2019cephalometric}  has shown commendable performance in the task of landmark detection.

\begin{figure}
\includegraphics[height=180px,keepaspectratio]{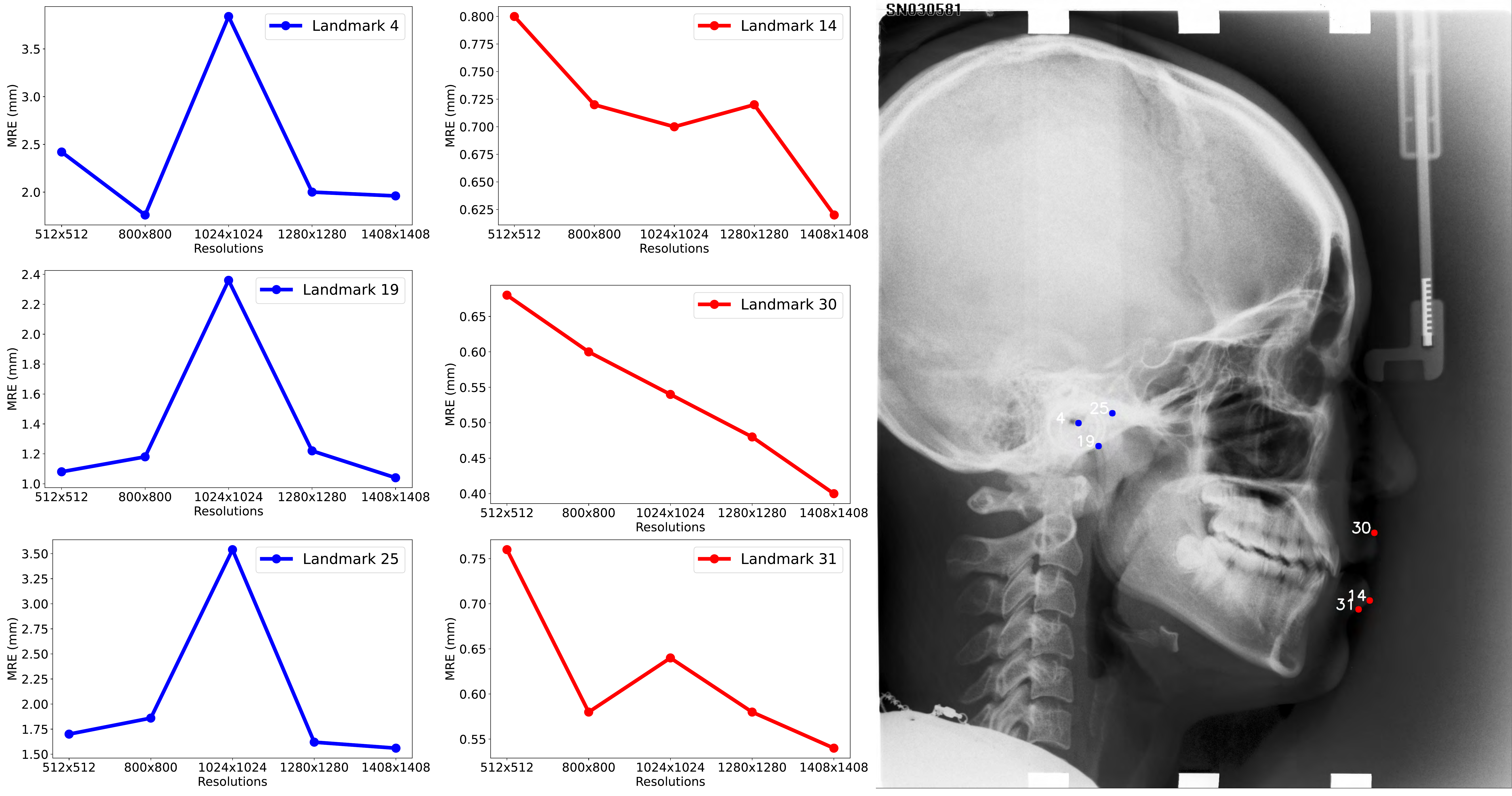}
\caption{The radial error of two groups of closely located landmarks at different resolutions.} 
\label{fig1}
\end{figure}

In this paper, we introduce an end-to-end method for cephalometric landmark detection that integrates multiple resolutions. Through experiments at different resolutions, we found that the same landmark has varying radial errors across different resolutions, and the distribution of radial errors for landmarks in different regions also differs depending on the resolution. (see Fig.~\ref{fig1})
The performance of the closely located landmarks 4, 19, and 25 at different resolutions shows a similar distribution, with smaller radial errors at both higher and lower resolutions. On the other hand, another group of closely located landmarks 14, 30, and 31, exhibit a completely different distribution across resolutions, with higher resolutions leading to smaller recognition errors. Therefore, we have reason to believe that landmarks in different positions have different preferences for the features extracted from various resolutions. Utilizing predictions from different resolutions can effectively reduce errors and enhance recognition accuracy.

To facilitate a comparison of our method with current popular techniques, we applied it to the cephalometric Landmark Detection in Lateral X-ray Images 2023 Challenge. This challenge aims to provide a comprehensive benchmark for the task of cephalometric landmark detection. In the final test set, our method achieved a SDR of 74.18\% and a MRE of 1.62.

\section{Methods}
\begin{figure}
\includegraphics[width=\textwidth]{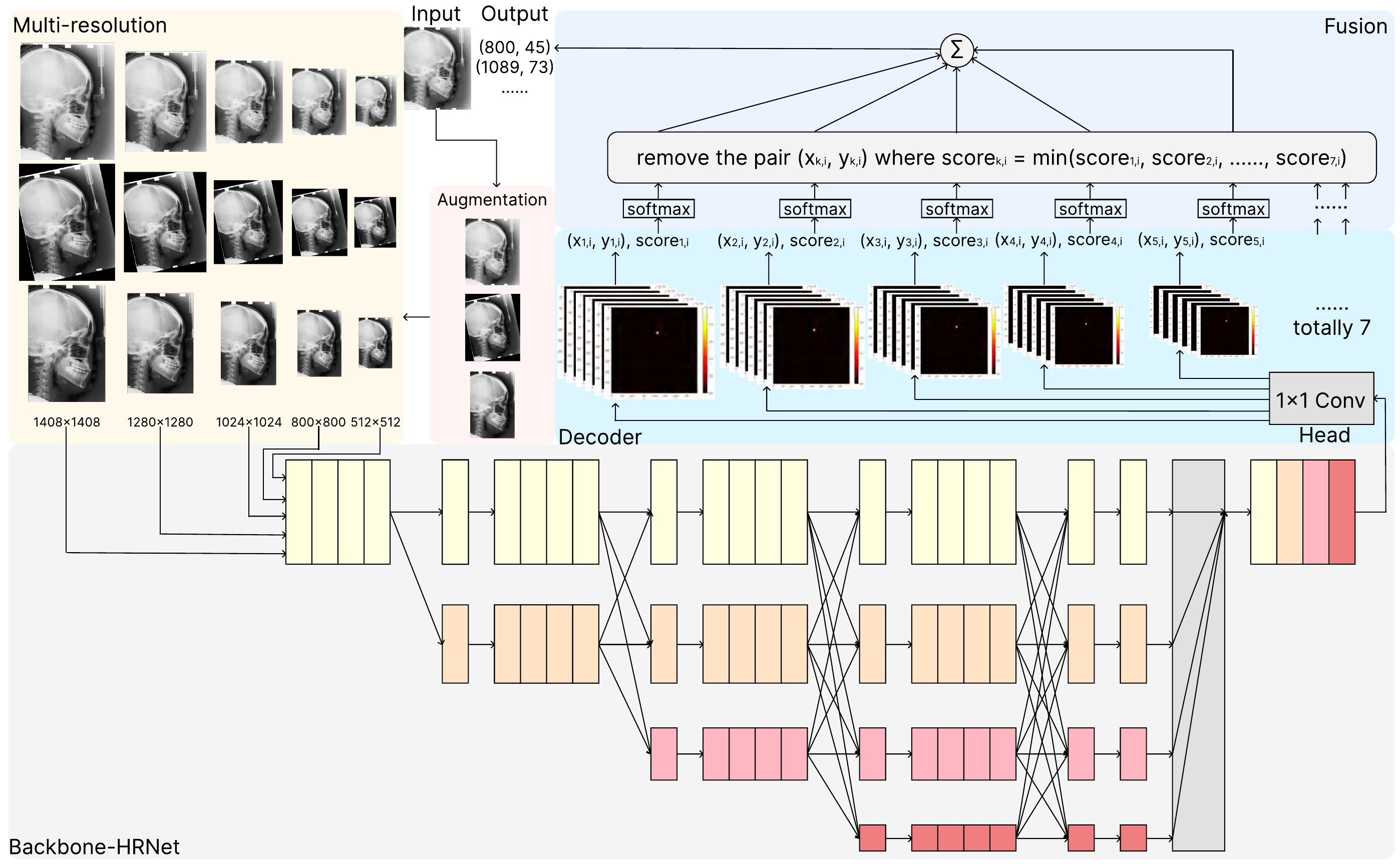}
\caption{The architecture of our method. The first step is to perform data augmentation, the second step is to resize the augmented images to five different resolutions, the third step is to individually input the images of different resolutions into the backbone network, the fourth step is to generate 38 heatmaps through the head module and convert the heatmaps into specific coordinates via a decoder, and the fifth step is fusion.} 
\label{fig2}
\end{figure}

The key of the cephalometric landmark detection is to, given a cephalometric X-ray image, detect landmarks in the image and obtain their two-dimensional coordinates. We utilized the baseline model provided officially for the competition, which was constructed using the MMPose framework \cite{mmpose2020}. We retained most of the structure from the baseline model, including using HRNet \cite{wang2020deep} as the backbone and the structure of the head module. Based on this, we made our own enhancements and optimizations. In essence, our proposed method's core is to train several models within the same framework using images of different resolutions as input. These models predict independently, and the final prediction results are filtered based on their respective confidence.(see Fig.~\ref{fig2})

\subsection{Multi Resolution}
Among the 38 landmarks in the dataset, landmarks at different locations have varying preferences for features. Some landmarks require higher resolution and detailed structural information, implying that a higher-resolution input is needed for the prediction. In contrast, some landmarks of the X-ray image demand deeper semantic information, suggesting the use of a lower-resolution input for prediction. Hence, to cater to the unique features of different landmarks, we employed various resolutions as input.

Specifically, we experimented with five resolutions: \( 512 \times 512 \), \( 800 \times 800 \), \( 1024 \times 1024 \), \( 1280 \times 1280 \), and \( 1408 \times 1408 \). The experimental results for each resolution can be found in Section 3 of the article.

\subsection{Data augmentation}
\begin{figure}
\includegraphics[width=\textwidth]{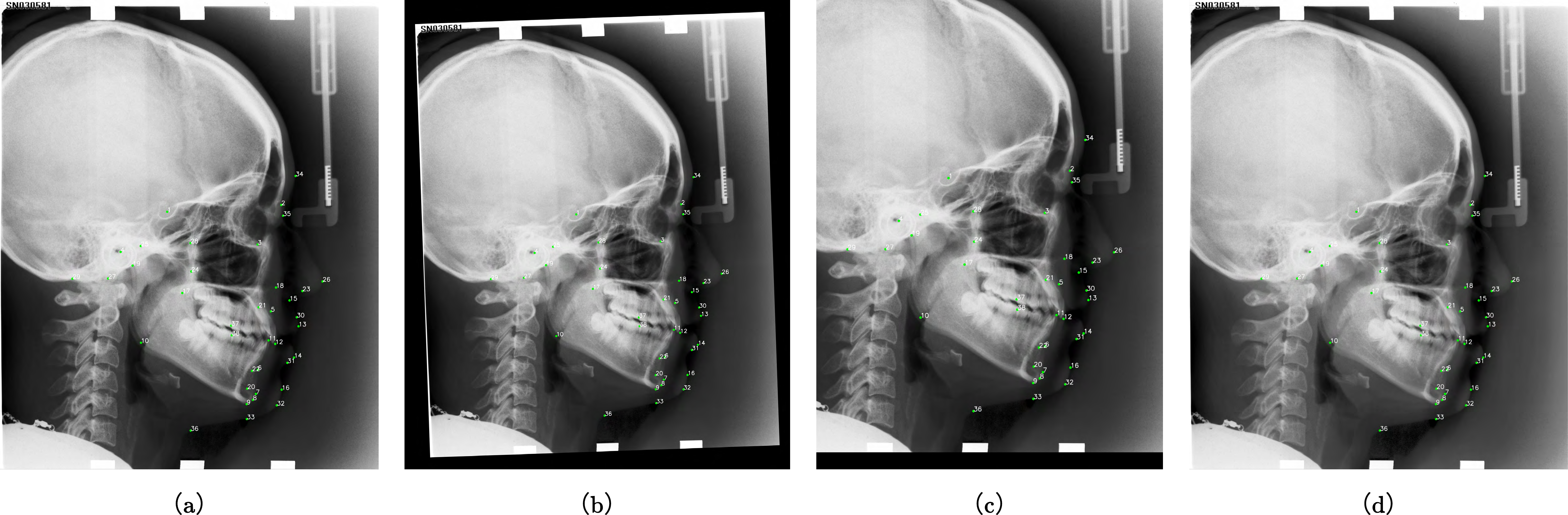}
\caption{The visualization of the data augmentations. (a) is the original image, (b) is the image after applying random rotation, (c) is the image after applying random shifting and cropping, (d) is the image after adjusting brightness.} 
\label{fig3}
\end{figure}

In the 400 X-ray images we obtained, due to variations in the tilt angle of patients' heads during imaging and misalignment during X-ray film scanning, some images appeared as if they had undergone a rotation. As a result, we trained a separate model with an input image resolution of \( 800 \times 800 \), incorporating random rotation(see Fig.~\ref{fig3}(b)) for data augmentation.

Owing to differences in patients' ages and imaging equipments, some images did not display a complete skull. Therefore, we trained five models with input image resolutions of \( 800 \times 800 \), \( 1024 \times 1024 \), \( 1280 \times 1280 \), and \( 1408 \times 1408 \) (two models at \( 800 \times 800 \)), incorporating random cropping and shifting(see Fig.~\ref{fig3}(c)) for data augmentation. Before implementing random rotation, cropping, and shifting, we padded 100 pixels of the borders of the images, enlarging the original size to prevent landmarks from being shifted out of the image during the augmentation process.

Regarding the imaging principle of X-ray photographs, denser bone tissue absorbs more X-rays, appearing white or pale gray in images. In contrast, the less dense soft tissue structure is penetrated by X-rays, manifesting as dark gray or even black in the images. Moreover, due to variations in dosage and exposure duration utilized by different devices and medical institutions, there's a notable brightness disparity across images. Thus, we trained a model with random brightness augmentation at a resolution of \( 512 \times 512 \). Adjusting the brightness of some images can make the contours of soft tissue structures more prominent(see Fig.~\ref{fig3}(d)), enhancing the identification of landmarks located at the edges of the soft tissue.

In summary, we trained a total of seven models with varying resolutions and data augmentation techniques for prediction.

\subsection{Result fusion}
In the head module, we adopted a heatmap-based prediction method. Our initial fusion strategy was to select based on the predictions of each model for each landmark on a local test set. However, as this approach was strongly tied to the local test set, its performance during the online testing phase was suboptimal. Subsequently, we made selections based on the confidence obtained during the model's predictions for each landmark. Among the seven prediction results for each landmark, we discarded the one with the lowest confidence. The remaining six prediction results were then averaged to obtain the final predicted coordinates. Since the confidence calculated by the model's decoder module were not normalized, we passed each confidence level through a softmax function for normalization before discarding the prediction with the lowest confidence.

\section{Evaluations and Discussions}
\subsection{Dataset}
The competition organizers provided a dataset including 600 X-ray images. These images come from three different medical institutions, hence have different sizes, with spacing values of 0.1, 0.125, and 0.096 respectively. Each image contains 38 landmarks. These landmarks were annotated separately by two medical experts, and the ground truth coordinates were averaged from the annotations of the two doctors. The image resolutions are \(2880 \times 2304 \), \(2089 \times 1937 \), and \(1935 \times 2400 \) respectively. To distribute different-sized images uniformly, they were padded according to the largest image and packaged as MHA files. Some of the images in the dataset were obtained by directly scanning existing X-ray films. Therefore, how to handle these images with relatively low clarity is one of the challenges of this competition. A high-quality dataset has significant implications for research. Compared to the challenges held at IEEE ISBI in 2014 and 2015 \cite{wang2015evaluation,wang2016benchmark}, the dataset provided by the organizers of this event has seen improvements both in terms of scale and the number of landmarks.

During the competition, 400 images were initially send to the participants for training, followed by 50 images for the online validation phase and 150 images for the testing phase. The 200 images used in the validation and testing phases were kept confidential throughout the competition. The ranking of the participants was based on their performance in the final testing phase.

\subsection{Analysis}
When using the acquired 400 images for local training, we randomly split them into 300, 50, and 50 for the training set, validation set, and test set, respectively. The final model submitted for online testing was trained on a combined dataset of the training and validation sets, using 350 images. To achieve better results in the competition, we conducted various experiments with different resolutions and data augmentation methods. The HRNet \cite{wang2020deep} backbone network used for feature extraction was pre-trained on the ImageNet dataset \cite{krizhevsky2012imagenet}. We adopted the MMPose framework \cite{mmpose2020} based on PyTorch for our implementation. The training process utilized the Adam optimizer \cite{kingma2014adam} with an initial learning rate of 0.0005, and we employed the default parameters of LinearLR and MultiStepLR as learning rate schedulers. For the model with a resolution of \(512 \times 512 \), training was conducted on a GTX 2080 TI GPU with a batch size of 8 and 250 epochs, which took approximately 2.5 hours. For higher resolutions, we utilized the GTX 4090 TI GPU for training. Depending on the resolution, batch sizes were set between 2-8. Training ran for 250 epochs, averaging around 5 hours.

When comparing our proposed method with the competition's baseline model on our local test set, the MRE improved by 0.6mm, and the SDR 2.0mm increased by 12.16\%. When tested on the online test set, the MRE improved by 0.36mm, and the SDR 2.0mm increased by 10.08\%.

\begin{table}
    \centering
    \caption{Comparison of the performance with the baseline provided by the official for both the local test dataset and the online test dataset.}
    \label{table1}
    
    \begin{tabular}{l|*{6}{p{1.3cm}|}p{1.3cm}}
    \hline
    \multirow{2}{*}{Model} & \multicolumn{5}{c|}{Local Test Dataset} & \multicolumn{2}{c}{Online Test Dataset} \\
    \cline{2-8}
    & \centering MRE & \centering 2 mm & \centering 2.5 mm & \centering 3 mm & \centering 4 mm & \centering MRE & \centering 2 mm \tabularnewline
    \hline
    \hline
    Baseline & \centering 2.09 & \centering 66.16 & \centering 76.79 & \centering 83.95	& \centering 91.90 & \centering 1.98 & \centering 64.10 \tabularnewline
    \hline
    \textbf{Ours} & \centering \textbf{1.49} & \centering \textbf{78.32} & \centering \textbf{84.90} & \centering \textbf{88.68}	& \centering \textbf{94.37} & \centering \textbf{1.62} & \centering \textbf{74.18} \tabularnewline
    \hline
    \end{tabular}
    
\end{table}

\begin{figure}
\includegraphics[width=\textwidth]{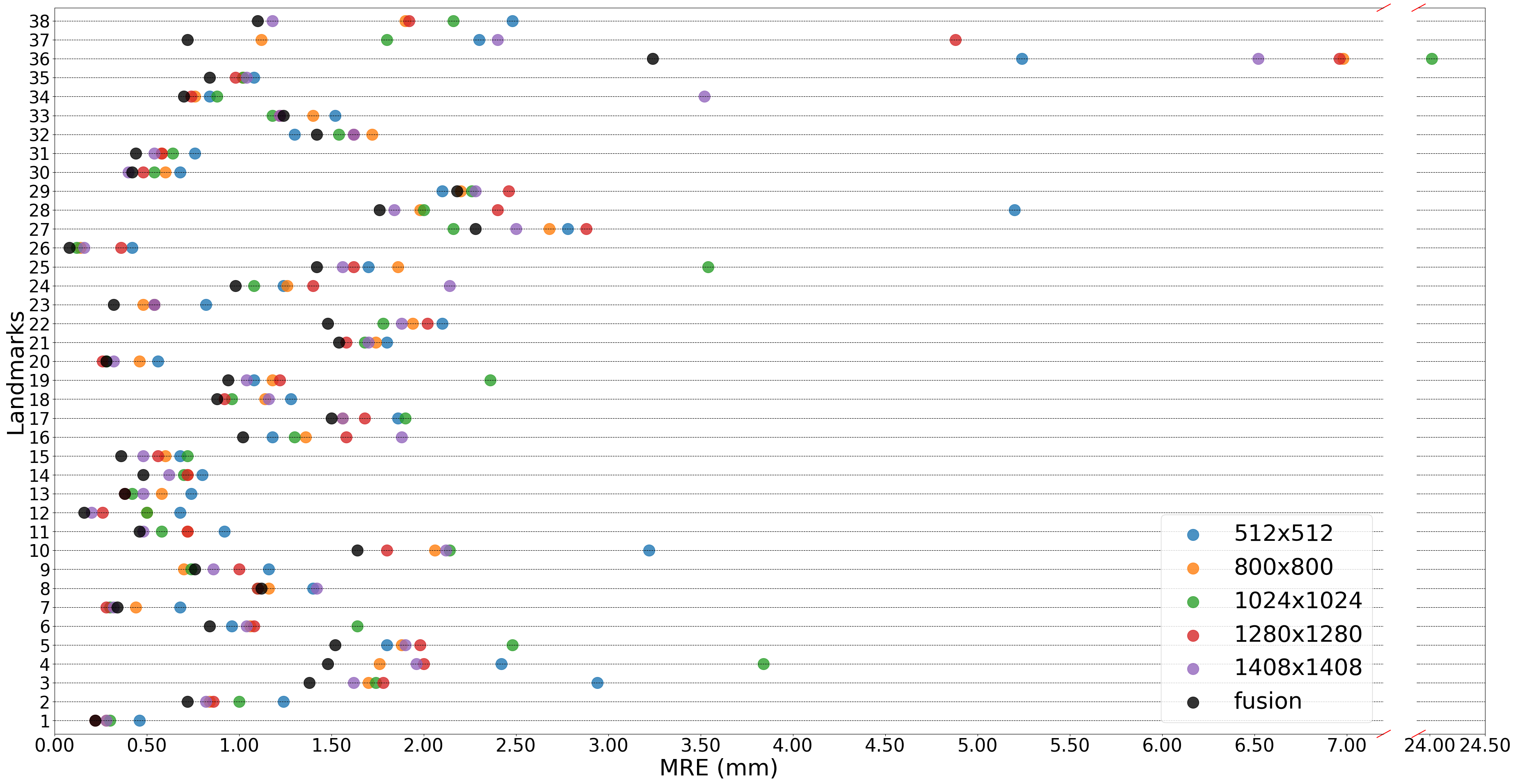}
\caption{The performance of each landmark at different resolutions.} 
\label{figure4}
\end{figure}

The performance disparities across different resolutions are more pronounced at specific landmarks. We set different resolutions as the input to the model, with the overall prediction results shown in Table~\ref{table2}., and the variation between each landmark is illustrated in Figure~\ref{figure4}. In Table~\ref{table2}, we did not use any data augmentation, shows that both overly small and overly large resolutions lead to decreased performance. The possible reasons could be that a resolution that's too small prevents the model from capturing the detailed information in the image, while an excessively large resolution results in insufficient extraction of deeper semantic information. Generally speaking, higher resolutions achieve better performance. As observed from Figure~\ref{figure4}, the model's recognition performance for landmark 36 remains suboptimal, and the radial error significantly decreases after fusion(the black points in the figure).

\begin{table}
    \centering
    \caption{Comparison of the performance with the different resolutions.(without augmentation)}
    \label{table2}
    
    \begin{tabular}{p{2cm}|>{\centering\arraybackslash}p{1.5cm}|>{\centering\arraybackslash}p{1.5cm}|>{\centering\arraybackslash}p{1.5cm}|>{\centering\arraybackslash}p{1.5cm}|>{\centering\arraybackslash}p{1.5cm}}
    \hline
    Resolution & \centering MRE & \centering 2 mm & \centering 2.5 mm & \centering 3 mm & \centering 4 mm \tabularnewline
    \hline
    \hline
    \(256 \times 256 \) & 2.42 & 50.16 & 63.58 & 74.00 & 86.84 \\
    \hline
    \(512 \times 512 \) & 2.09 & 69.42 & 79.68 & 85.42 & 92.16 \\
    \hline
    \(800 \times 800 \) & 1.84 & 72.32 & 80.26 & 85.74 & 92.63 \\
    \hline
    \(1024 \times 1024 \) & 1.87 & 71.21 & 79.21 & 84.05 & 90.84 \\
    \hline
    \(1280 \times 1280 \) & 1.94 & 73.63 & 80.01 & 84.94 & 90.58 \\
    \hline
    \(1408 \times 1408 \) & 1.90 & 74.31 & 81.95 & 86.95 & 92.32 \\
    \hline
    \(2048 \times 2048 \) & 4.70 & 67.95 & 74.63 & 79.58 & 86.53 \\
    \hline
    \end{tabular}
    
\end{table}

Similarly, we compared the performance before and after data augmentation, with the results shown in Table~\ref{table3}. It is evident that randomly cropping and shifting the images significantly improves prediction accuracy. Among the four different resolutions tested, the augmented SDR 2.0mm can exceed 75\%, demonstrating that the data augmentation we performed is indeed effective.

\begin{table}
    \centering
    \caption{Comparison of the performance before and after data augmentation.}
    \label{table3}
    
    \begin{tabular}{l|l|>{\centering\arraybackslash}p{1.3cm}|>{\centering\arraybackslash}p{1.3cm}|>{\centering\arraybackslash}p{1.3cm}|>{\centering\arraybackslash}p{1.3cm}|>{\centering\arraybackslash}p{1.3cm}}
    \hline
    Resolution & Augmentation & \centering MRE & \centering 2 mm & \centering 2.5 mm & \centering 3 mm & \centering 4 mm \tabularnewline
    \hline
    \hline
    \multirow{2}{*}{\(512 \times 512 \)} & Nothing & 2.09 &	69.42 &	79.68 &	85.42 & 92.16 \\
    \cline{2-7}
    & Adjust brightness & 1.92 & 66.53 & 76.42 & 83.00 & 91.26 \\
    \hline
    \multirow{3}{*}{\(800 \times 800 \)} & Nothing & 1.84 &	72.32 &	80.26 &	85.74 &	92.63 \\
    \cline{2-7}
    & Shifting \& cropping & 1.80 & 74.42 & 81.73 & 86.37 & 92.16 \\
    \cline{2-7}
    & Rotation & 1.60 &	75.90 &	82.79 &	87.58 &	94.00 \\
    \hline
    \multirow{2}{*}{\(1024 \times 1024 \)} & Nothing & 1.87 & 71.21 & 79.21 & 84.05 & 90.84 \\
    \cline{2-7}
    & Shifting \& cropping & 1.71 &	75.32 &	81.58 &	86.32 &	92.16 \\
    \hline
    \multirow{2}{*}{\(1280 \times 1280 \)} & Nothing & 1.94 & 73.63 & 80.01 & 84.94 & 90.58 \\
    \cline{2-7}
    & Shifting \& cropping & 1.88 &	75.53 &	82.21 &	86.37 &	91.90 \\
    \hline
    \multirow{2}{*}{\(1408 \times 1408 \)} & Nothing & 1.90 & 74.31 & 81.95 & 86.95 & 92.32 \\
    \cline{2-7}
    & Shifting \& cropping & 1.83 &	75.47 &	82.05 &	86.90 &	92.47 \\
    \hline
    \end{tabular}
    
\end{table}

\section{Conclusion}
In this article, we introduce a method for cephalometric landmark detection that integrates multiple resolutions, significantly improving model performance through a variety of straightforward data augmentations. In our approach, the diverse resolutions ensure that landmarks at different positions can utilize semantic features of varying levels for prediction. Given the characteristic of limited medical imaging data, we believe that more effective data augmentation strategies can be added in the future. For instance, employing image registration algorithms, selecting specific reference landmarks for alignment, or utilizing generative models to augment the dataset could be beneficial. Furthermore, the recognition task can be divided into two stages, with a more precise identification being conducted within the Region of Interest (ROI).

\bibliographystyle{splncs04}
\bibliography{references}

\end{document}